\documentclass[12pt]{article}
\usepackage[utf8]{inputenc}
\usepackage{amsmath}
\usepackage{amsfonts}
\usepackage{amssymb}
\usepackage{bm}
\usepackage{amsthm}
\usepackage{slashbox}
\usepackage{array,multirow,makecell}
\usepackage{graphicx}
\usepackage{multicol}
\usepackage{subcaption}
\usepackage{mdwlist}
\setcellgapes{1pt}
\makegapedcells
\newcolumntype{R}[1]{>{\raggedleft\arraybackslash }b{#1}}
\newcolumntype{L}[1]{>{\raggedright\arraybackslash }b{#1}}
\newcolumntype{C}[1]{>{\centering\arraybackslash }b{#1}}

\begin{document}


\begin{center}
{\Large
	{\sc  R\'eint\'egration des refus\'es en \textit{Credit Scoring}}
}
\medskip

 A. Ehrhardt $^{1}$ \& C. Biernacki $^{2}$ \& V. Vandewalle $^{3}$ \& P. Heinrich $^{4}$ \& S. Beben $^{5}$
\medskip

{\it
$^{1}$ Crédit Agricole Consumer Finance, Inria, aehrhardt@ca-cf.fr
 
$^{2}$ Inria, Université de Lille 1, christophe.biernacki@{univ-lille1,inria}.fr

$^{3}$ Inria, Université de Lille 2, vincent.vandewalle@{univ-lille2,inria}.fr

$^{4}$ Université de Lille 1 Laboratoire Paul Painlevé, philippe.heinrich@univ-lille1.fr

$^{5}$ Crédit Agricole Consumer Finance, sebeben@ca-cf.fr
}
\end{center}
\smallskip


{\bf R\'esum\'e.} Un système d’octroi de crédit peut refuser des demandes de prêt jugées trop risquées. Au sein de ce système, le score de crédit fournit une valeur mesurant un risque de défaut, valeur qui est comparée à un seuil d'acceptabilité. Ce score est construit exclusivement sur des données de clients financés, contenant en particulier l'information ``bon ou mauvais payeur'', alors qu'il est par la suite appliqué à l'ensemble des demandes.  Un tel score est-il statistiquement pertinent ? Dans cette note, nous précisons et formalisons cette question et  étudions l’effet de l’absence des non-financés sur les scores élaborés. Nous présentons ensuite des méthodes pour réintégrer les non-financés et concluons sur leur inefficacité en pratique, à partir de données issues de Crédit Agricole Consumer Finance.

{\bf Mots-cl\'es.} r\'eint\'egration, refus\'e, scoring, risque, cr\'edit, apprentissage semi-supervis\'e
\smallskip

{\bf Abstract.} The granting process of all credit institutions rejects applicants who seem risky regarding the repayment of their debt. A credit score is calculated and associated with a cut-off value beneath which an applicant is rejected. Developing a new score implies having a learning dataset in which the response variable good/bad borrower is known, so that rejects are \textit{de facto} excluded from the learning process. We first introduce the context and some useful notations. Then we formalize if this particular sampling has consequences on the score's relevance. Finally, we elaborate on methods that use not-financed clients' characteristics and conclude that none of these methods are satisfactory in practice using data from Crédit Agricole Consumer Finance. 



\section{Introduction}

L'objectif d'une institution financière est de savoir si un demandeur de crédit est susceptible de faire défaut lorsqu'on lui accorde un prêt.
On dispose de $n$ clients financés possédant $d$ caractéristiques et que l'on note $\bm{x}^{\text{f}} = \{x_i \in \mathbb{R}^d : 1\leq i \leq n\}$ et $m$ clients non financés dont les $d$ caractéristiques sont $\bm{x}^{\text{nf}} = \{x_i : n+1\leq i \leq n+m\}$. On définit ${\bm{y}}^{\text{f}}$ et ${\bm{y}}^{\text{nf}}$ de manière équivalente pour désigner la variable cible binaire ``bon ou mauvais payeur''.
On se place dans le cadre des modèles de régression logistique $\{p_{\theta}(y|x)\}_{\theta \in \Theta}$, standard dans le milieu du crédit, et on cherche à estimer $\theta$. Une approche classique est de maximiser la log-vraisemblance complète suivante, conduisant à l'estimateur $\hat{\theta}$ :
\[\ell(\theta;{\bm{x}},{\bm{y}}) = \sum_{i=1}^{n} \ln(p_{\theta}(y_i|x_i)) + \sum_{i=n+1}^{n+m} \ln(p_{\theta}(y_i|x_i)) = \ell(\theta;{\bm{x}}^{\text{f}},{\bm{y}}^{\text{f}}) + \ell(\theta;{\bm{x}}^{\text{nf}},{\bm{y}}^{\text{nf}}).\]
Malheureusement, ${\bm{y}}^{\text{nf}}$ est inconnu : on ne sait pas si un client non financé aurait remboursé son crédit ; on ne peut donc pas calculer $\ell(\theta;{\bm{x}}^{\text{nf}},{\bm{y}}^{\text{nf}})$. On maximise alors traditionnellement la vraisemblance observée $\ell(\theta;{\bm{x}}^{\text{f}},{\bm{y}}^{\text{f}})$ et on obtient $\hat{\theta}^{\text{f}}$.

On a donc d'une part un estimateur ``idéal'' $\hat{\theta}$ qu'on ne peut obtenir en pratique. D'autre part, on obtient l'estimateur $\hat{\theta}^{\text{f}}$ par la méthode actuelle de construction des scores.
Il convient dans un premier temps d'étudier l'impact de la sélection des clients financés sur ces deux estimateurs. On s'intéresse ensuite aux méthodes existantes visant à réduire cet impact.

\section{Impacts des clients refusés sur les estimateurs}

\paragraph*{Asymptotiques des deux estimateurs}

Sous certaines hypothèses de régularité, on sait d'après White (1982) qu'il existe $\theta_{\text{opt}}$, $\theta_{\text{opt}}^{\text{f}} \: \in \: \Theta$ tels que $\hat{\theta} \xrightarrow[n,m \to \infty]{a.s.} \theta_{\text{opt}}$ et $\hat{\theta}^{\text{f}} \xrightarrow[n \to \infty]{a.s.} \theta_{\text{opt}}^{\text{f}}$. De plus, $\sqrt[]{n+m} ( \hat{\theta} - \theta_{\text{opt}} ) \xrightarrow[n,m \to \infty]{\mathcal{L}} \mathcal{N}_{d+1} (0,{\Sigma}_{\theta_{\text{opt}}})$ et $ \sqrt[]{n} ( \hat{\theta}^{\text{f}} - \theta_{\text{opt}}^{\text{f}} ) \xrightarrow[n \to \infty]{\mathcal{L}} \mathcal{N}_{d+1} (0,{\Sigma}^{\text{f}}_{\theta_{\text{opt}}^{\text{f}}})$ où ${\Sigma}_{\theta_{\text{opt}}}$ and  ${\Sigma}^{\text{f}}_{\theta_{\text{opt}}^{\text{f}}}$ sont les matrices de variance-covariance asymptotiques de ces deux estimateurs, liées à leur matrice d'information de Fisher.

La question de l'impact d'un score de risque de crédit appris sur une population de financés plutôt qu'un score appris sur toute la population peut donc être divisée en deux sous-questions :
\begin{enumerate*}
\item[Q1.] Est-ce que les scores sont asymptotiquement (en $n,m$) les mêmes, i.e. $\theta_{\text{opt}} = \theta^{\text{f}}_{\text{opt}}$?
\item[Q2.] Les matrices de variance-covariance asymptotiques sont-elles égales : ${\Sigma}_{\theta_{\text{opt}}} = {\Sigma}^{\text{f}}_{\theta_{\text{opt}}^{\text{f}}}$?
\end{enumerate*}

Pour répondre à (Q1) et (Q2), on discute deux hypothèses : la première est celle de la validité du modèle logistique, la seconde porte sur le mécanisme des données manquantes.

\paragraph*{Validité du modèle logistique}
Lorsqu'on choisit un modèle, on fait souvent l'hypothèse implicite que celui-ci est bien spécifié, c'est-à-dire que les données proviennent effectivement de la distribution choisie lors de la modélisation: il existe $\theta_{\text{vrai}} \in \Theta$ tel que $p(y|x)=p_{\theta_{\text{vrai}}}(y|x)$.
Cette hypothèse est rarement vérifiée en pratique mais on essaie de s'en approcher par l'utilisation de modèles robustes (comme la régression logistique) et la discrétisation des variables continues (voir Tufféry (2010) pour l'intérêt de la discrétisation des variables en régression logistique).

\paragraph*{Mécanisme des données manquantes}
Le mécanisme de sélection génère des valeurs manquantes $\bm{y}^{\text{nf}}$. La probabilité d'être financé sachant $x$ et $y$, $p(\text{f}|x,y)$, peut être caractérisée de trois façons introduites par Little et Rubin (2014): MCAR, MAR et MNAR. Le cas MCAR (complétement au hasard) est écarté dans la suite. Le cas MAR correspond à une sélection dépendant seulement de $x$, i.e. $p(\text{f}| x,y) = p(\text{f}| x)$, cette hypothèse serait ainsi vraie si l'acceptation était basée uniquement sur le score par exemple. Le cas MNAR où $p(\text{f}| x,y) \neq p(\text{f}| x)$ est l'hypothèse la plus plausible en \textit{Scoring} dans la mesure où les conseillers clientèle peuvent exploiter des variables prédictives pertinentes non intégrées à $x$ pour appuyer leurs décisions de financement.

L'étude de l'influence de ces deux types d'hypothèse sur $\hat{\theta}$ et $\hat{\theta}^{\text{f}}$ permet de répondre aux questions (Q1) et (Q2) dans le tableau~\ref{tableasymptotic}. Ces réponses sont négatives dans la plupart des cas, d'où la nécessité de ``corriger'' $\hat{\theta}^{\text{f}}$ grâce à différents moyens d'action exploités par les méthodes que nous décrivons brièvement dans la partie suivante.

\begin{table}[htbp]
\centering{\begin{tabular}{|R{4.5cm}||C{4.5cm}|C{4.5cm}|}
\hline & MAR & MNAR \\
 & $\forall~x,y, \: p(\text{f}| x,y) = p(\text{f}| x)$ & $\exists \: x,y, \: p(\text{f}| x,y) \neq p(\text{f}| x)$ \\
\hline
\hline Vrai modèle atteint & $\theta_{\text{opt}}^{\text{f}} = \theta_{\text{opt}}$ &  \\ 
$p(y|x)\in \{p_\theta(y|x)\}_{\theta\in\Theta}$  & ${\Sigma}^{\text{f}}_{\theta_{\text{opt}}^{\text{f}}} \neq {\Sigma}_{\theta_{\text{opt}}}$ & $\theta_{\text{opt}}^{\text{f}} \neq \theta_{\text{opt}}$ \\ \cline{1-2}
 Vrai modèle non atteint & $\theta_{\text{opt}}^{\text{f}} \neq \theta_{\text{opt}}$ & ${\Sigma}^{\text{f}}_{\theta_{\text{opt}}^{\text{f}}} \neq {\Sigma}_{\theta_{\text{opt}}}$ \\
$p(y|x)\notin\{p_\theta(y|x)\}_{\theta\in\Theta}$ & ${\Sigma}^{\text{f}}_{\theta_{\text{opt}}^{\text{f}}} \neq {\Sigma}_{\theta_{\text{opt}}}$ &  \\
\hline 
\end{tabular}}
\caption{Egalité de $\theta_{\text{opt}}$ et $\theta_{\text{opt}}^{\text{f}}$ (Q1) et celle de ${\Sigma}_{\theta_{\text{opt}}}$ et ${\Sigma}^{\text{f}}_{\theta_{\text{opt}}^{\text{f}}}$ (Q2) selon les hypothèses sur $p_{\theta}(y|x)$ et $p(\text{f}|x,y)$.}
\label{tableasymptotic}
\end{table}

\vspace{-1em}

\section{Méthodes dites de ``réintégration des refusés''}

Un organisme de crédit souhaiterait améliorer $\hat{\theta}^{\text{f}}$ au sens de (Q1) et (Q2) du tableau~\ref{tableasymptotic}. Pour ce faire, il faudrait soit changer la famille de modèles $\Theta$ proposée, ce que l'on s'interdit car la régression logistique est imposée, soit modéliser le mécanisme d'acceptation $p(\text{f}|x,y)$, soit enfin utiliser l'information disponible mais non encore utilisée $\bm{x}^{\text{nf}}$.

Ces deux derniers moyens d'action sont utilisés indirectement par plusieurs auteurs, notamment Feelders (2000), Viennet et al. (2006), Banasik et Crook (2007), Guizani et al. (2013) et Nguyen (2016), qui proposent des méthodes empiriques d'estimation de $\theta$ utilisant $\bm{x}^{\text{nf}}$ et faisant parfois implicitement des hypothèses supplémentaires sur le mécanisme des données manquantes. Nous nous concentrons ici sur trois méthodes~: deux issues de la litérature dédiée, \textit{Augmentation} et \textit{Parceling}, ainsi que le modèle génératif sur $p(x,y)$.

On introduit le modèle génératif $\{p_{\alpha}(x,y)\}_{\alpha \in \text{A}}$ qui permet d'utiliser toute l'information disponible, de faire des prédictions (par la règle de Bayes), et donc d'être considérée comme une méthode de ``réintégration des refusés''. L'utilisation de $\bm{x}^{\text{nf}}$ est permise par la modélisation de $p(x)$ et l'utilisation de l'algorithme EM pour l'estimation des paramètres $\alpha \in \text{A}$. Cette méthode simple est adaptée au cas MAR. Sous l'hypothèse du vrai modèle, elle permet d'obtenir une efficacité asymptotique (Q2) meilleure que $\hat{\theta}^{\text{f}}$ (voir notamment O'neill (1980)). En revanche, dans le cas contraire, ce modèle est potentiellement sujet à plus de biais (Q1) qu'un modèle prédictif du fait de la modélisation de $p(x)$.

La méthode d'\textit{Augmentation} propose de constituer des ``bandes de score'', i.e. des regroupements de clients à la probabilité de défaut $p_{\hat{\theta}^{\text{f}}}(y|x)$ ``proche'' (ce qui suppose l'estimation de $\hat{\theta}^{\text{f}}$ en premier lieu) et de pondérer les observations de clients financés de chaque bande de score respectivement par l'inverse de la proportion de clients financés au sein de la bande (voir notamment Banasik et Crook (2007)). On peut montrer que, dans le cas MAR et vrai modèle non atteint et en supposant $p(\text{f}|x) >0$ quel que soit $x$, une pondération des observations de $\ell(\theta;{\bm{x}}^{\text{f}},{\bm{y}}^{\text{f}})$ par $1/p(\text{f}|x)$ conduit à un estimateur $\hat{\theta}^{\text{f}}$ égal à $\hat{\theta}$, permettant ainsi de répondre à (Q1) par l'affirmative. Ne connaissant pas $p(\text{f}|x)$, cette méthode tente d'approcher cette correction ``idéale''. En pratique, outre les questions du choix du nombre de bandes de score et de la façon de les constituer, l'hypothèse $p(\text{f}|x) > 0$ n'est pas satisfaite : certaines régions de l'espace (étudiants, chômeurs, ...) n'étant jamais explorées.

La méthode de \textit{Parceling} propose, dans le cas MNAR, de constituer des bandes de score (clients à la probabilité de défaut $p_{\hat{\theta}^{\text{f}}}(y|x)$ ``proche'') et d'attribuer aux non-financés de chaque bande un taux de défaut majoré (paramètre à fixer par le statisticien) par rapport à celui des financés de la bande. On tire ensuite les observations manquantes $\bm{y}^{\text{nf}}$ selon le taux de défaut de la bande à laquelle elles appartiennent. On ré-entraîne enfin le modèle logistique (voir notamment Viennet et al. (2006)). Contrairement à l'\textit{Augmentation} qui ``corrigeait'' $\ell(\theta;{\bm{x}}^{\text{f}},{\bm{y}}^{\text{f}})$, le \textit{Parceling} propose de maximiser $\ell(\theta;{\bm{x}},{\bm{y}})$ en tirant ${\bm{y}}^{\text{nf}}$ selon une loi choisie par le statisticien. Ce choix ne relève pas d'un processus d'estimation : on fait des hypothèses invérifiables (notamment MNAR) dont la pertinence dans le modèle final ne peut être évaluée.

\begin{figure}[!h]
{\setlength{\parindent}{0cm}
\begin{subfigure}[t]{\textwidth}
\begin{center}
\includegraphics[width=14cm]{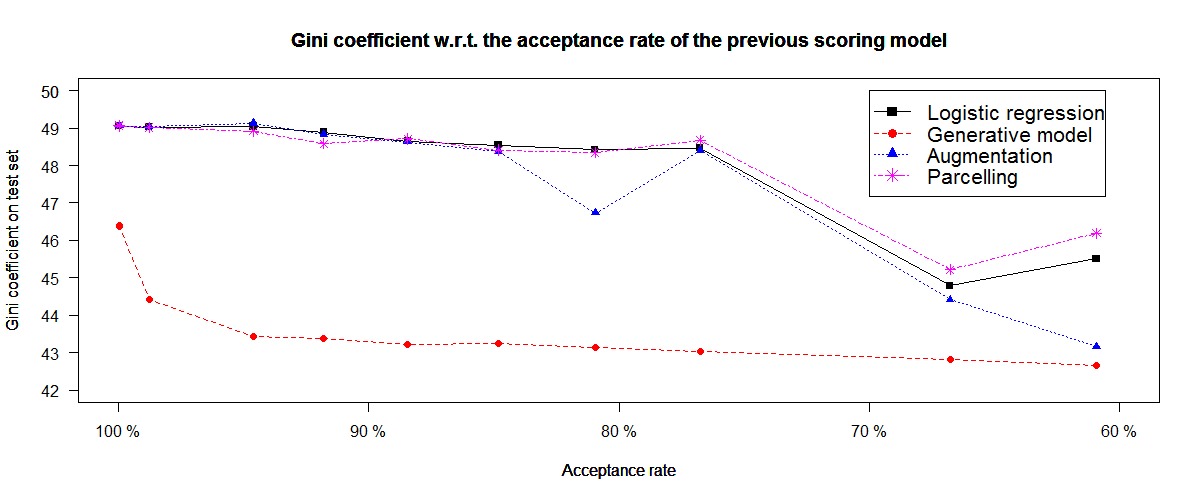}
\caption{Grande distribution d'articles de sport}
\label{fig:methods4}
\end{center}
\end{subfigure}

\begin{subfigure}[t]{\textwidth}
\begin{center}
\includegraphics[width=14cm]{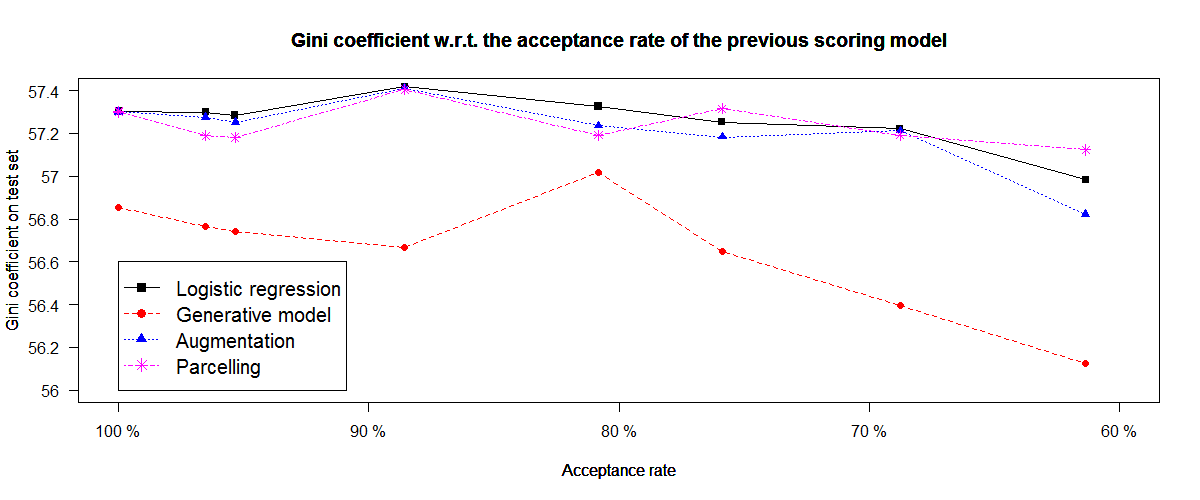}
\caption{Electroménager grand public}
\label{fig:rejectconsumer}
\end{center}
\end{subfigure}
}
\caption{Gini sur deux ensembles test des scores résultant de différentes méthodes de ``réintégration des refusés'' en fonction du taux d'acceptation}
\label{twins}
\end{figure}

Nous nous proposons de tester ces modèles expérimentalement et les comparer au modèle $\hat{\theta}^{\text{f}}$ pour deux jeux de données distincts provenant du Crédit Agricole Consumer Finance, l'un issu de la grande distribution d'articles de sport, l'autre issu de l'électroménager grand public, et pour lesquels on simule une sélection graduellement plus stricte. En pratique, les limitations des trois méthodes que nous avons évoquées précédemment se vérifient. On peut voir sur les figures~\ref{fig:methods4} et~\ref{fig:rejectconsumer} que le modèle génératif produit un indice de Gini plus faible que les autres modèles (y compris celui issu de $\hat{\theta}^{\text{f}}$) quel que soit la proportion de clients non financés du fait du biais introduit par la modélisation de $p(x)$. Les méthodes d'\textit{Augmentation} et de \textit{Parceling} produisent des résultats très similaires à $\hat{\theta}^{\text{f}}$, leur Gini étant (non significativement) parfois supérieur, parfois inférieur à $\hat{\theta}^{\text{f}}$. On remarque également une forte dépendance des résultats au jeu de données utilisé. En conclusion, il n'y a pas de méthode uniformément et significativement meilleure à $\hat{\theta}^{\text{f}}$, raison pour laquelle nous ne préconisons pas la pratique de la ``réintégration des refusés''.

\section{Conclusion}

Nous avons formalisé un problème ancien du domaine du \textit{Scoring} à travers les questions (Q1) et (Q2). Nous avons vu, selon la validité du modèle logistique et le mécanisme des données manquantes, que la méthode actuelle de construction des scores produisant $\hat{\theta}^{\text{f}}$ est, au mieux moins efficace que l'estimateur idéal $\hat{\theta}$ impossible à obtenir en pratique (Q2), et au pire asymptotiquement biaisée (Q1).

Pressentant vraisemblablement les lacunes de ce score, et en l'absence de formalisation, plusieurs auteurs ont proposé des méthodes empiriques de réintégration des refusés tentant d'améliorer le processus d'estimation du score. Nous nous sommes concentrés ici sur les méthodes~\textit{Augmentation} et \textit{Parceling}.

L'étude approfondie des outils développés dans la première partie a permis de ré-interpréter ces méthodes : nous avons exhibé dans quel cadre sur le modèle $p_{\theta}(y|x)$ et sur le mécanisme des données manquantes ces méthodes opèrent, leurs hypothèses implicites sur tout ou partie de la loi jointe $p(x,y,\text{f})$, mais aussi leurs limitations pratiques.

A travers nos expériences, nous confirmons les limitations théoriques de ces méthodes. On comprend aussi aisément pourquoi la question de la ``réintégration des refusés'' a divisé les auteurs, Guizani (2013) et Nguyen (2016) concluant à l'efficacité et la nécessité de tels méthodes, tandis que Viennet et al. (2006) et Banasik et Crook (2007) se révèlent plus mitigés. En effet, le succès d'une méthode par rapport à une autre est fortement conditionné aux données (il n'y a pas de méthode uniformément meilleure) comme l'illustrent les deux exemples de la figure~\ref{twins}, ce qui était d'ailleurs déjà souligné par Viennet et al. (2006).

\section*{Bibliographie}

\noindent [1] Banasik, J. et Crook, J. (2007), Reject inference, Augmentation, and sample selection, {\it European Journal of Operational Research}, 183(3):1582-1594.

\noindent [2] Feelders, A. (2000), Credit Scoring and reject inference with mixture models, {\it International Journal of Intelligent Systems in Accounting, Finance \& Management}, 9(1)1:8.

\noindent [3] Guizani, A. Souissi, B., Ammou, S.B. et Saporta, G. (2013), Une comparaison de quatre techniques d'inférence des refusés dans le processus d'octroi de crédit, {\it 45èmes journées de statistiques}.

\noindent [4] Kiefer, N.M. et Larson, C.E. (2006), Specification and informational issues in credit scoring, {\it Available at SSRN 956628}.

\noindent [5] Little, R.J. et Rubin, D.B. (2014), {\it Statistical Analysis with missing data}, John Wiley \& Sons.

\noindent [6] Nguyen, H.T. (2016), Reject inference in application scorecards: evidence from France, technical report, University of Paris-West la Défense, EconomiX.

\noindent [7] O'neill, T.J. (1980), The general distribution of the error rate of a classification procedure with application to logistic regression discrimination, \textit{Journal of the American Statistical Association}, 75(369):154-160.

\noindent [8] Tufféry, S. (2010), {\it Data Mining et statistique décisionnelle : l'intelligence des données}, Editions Technip.

\noindent [9] Viennet, E., Soulié, F.F. et Rognier, B. (2006), Evaluation de techniques de traitement des refusés pour l'octroi de crédit, {\it ArXiv preprint cs/0607048}.

\noindent [10] White, H. (1982), Maximum likelihood estimation of misspecified models, {\it Econometrica: Journal of the Econometric Society}, pages 1-25.

\end{document}